# Probing Subcellular Nanostructure of Engineered Human Cardiomyocytes in 3D Tissue

Josh Javor[1*], Jourdan K. Ewoldt[2], Paige E. Cloonan[2], Anant Chopra[2], Rebeccah J. Luu[2], Guillaume Freychet[3], Mikhail Zhernenkov[3], Karl Ludwig[4,7], Jonathan G. Seidman[5], Christine E. Seidman[5], Christopher S. Chen[1,2] and David J. Bishop[1,2,4,6,7]

1. Department of Mechanical Engineering, Boston University, Boston, Massachusetts, 02215, USA
2. Department of Biomedical Engineering, Boston University, Boston, Massachusetts, 02215, USA
3. SMI Beamline 12-ID, Brookhaven National Laboratory, Upton, NY 11973, USA
4. Department of Physics, Boston University, Boston, Massachusetts, 02215, USA
5. Department of Genetics, Harvard Medical School, Boston, MA, 02215, USA
6. Department of Electrical Engineering, Boston University, Boston, Massachusetts, 02215, USA
7. Division of Materials Science, Boston University, Boston, Massachusetts, 02215, USA

ABSTRACT

The structural and functional maturation of human induced pluripotent stem cell-derived cardiomyocytes (hiPSC-CMs) is essential for application to pharmaceutical testing, disease modeling, and ultimately therapeutic use. Multicellular 3D-tissue platforms have improved functional maturation of hiPSC-CMs, but probing cardiac contractile properties remains challenging in a 3D environment, especially at depth and in live tissues. Using small angle X-ray scattering (SAXS) images, we show that hiPSC-CMs, matured and examined in a 3D environment, exhibit periodic spatial arrangement of the myofilament lattice, which has not been previously detected in hiPSC-CMs. Contractile force is found to correlate with both scattering intensity ($R^2$=0.44) and lattice spacing ($R^2$=0.46). Scattering intensity also correlates with lattice spacing ($R^2$=0.81), suggestive of lower noise in our structural measurement relative to the functional measurement. Notably, we observe decreased myofilament ordering in tissues with a myofilament mutation known to lead to hypertrophic cardiomyopathy (HCM). Our results highlight the progress of human cardiac tissue engineering and enable unprecedented study of structural maturation in hiPSC-CMs.

INTRODUCTION

Engineering of 3D human cardiac tissue has advanced significantly in the last decade, promoting multicellular organization and functional maturation of human induced pluripotent stem cell-derived cardiomyocytes (hiPSC-CMs)[1-3]. While this is promising for applications such as drug development[4], disease modeling[5,6], and therapeutic implants[7], hiPSC-CMs require further maturation and reproducibility[8,9]. This has inspired many new technologies, such as three-dimensional (3D) tissue platforms[1-3], active mechanical environments[10], tissue training protocols[1,2], and techniques to probe subcellular information within the 3D environment[11-15], all of which then inform future maturation protocols for cardiac tissue. Despite the critical importance of subcellular structure to cardiac tissue function and maturation, few tools exist to probe the nanostructural organization of hiPSC-CMs in a 3D environment, especially in live tissue.

In the engineering of tissues from hiPSC-CMs, complex 3D organization and accessibility to subcellular information are areas of significant traction. In the native heart, complex 3D organization is instrumental to cardiac function. *In vitro* models with an engineered 3D environment promote improvements to structure and function relative to their two-dimensional (2D) monolayer counterparts[16,17,18]. Specifically, 3D models have been shown to improve cell morphology, contraction ability, presence of intracellular adhesion structures, organization of myofibrils, mitochondria morphology, endoplasmic reticulum contents, and expression of cardiac differentiation markers[16]. In parallel, subcellular techniques have been developed to probe function, such as action potential[11,12], traction force[13], and metabolism[14], but most are limited to 2D and surface readouts.

The advances in 3D tissue culture require measurement techniques, which can probe depth and live dynamics. Electron microscopy requires invasive sample preparation, with chemical fixation, microtome slicing, staining, and a high vacuum environment[1,2,16]. Not only do these invasive methods have the potential to distort structural features, but the throughput is low and this technique is not compatible with live tissue imaging. Light-based methods fundamentally face the diffraction limit and cannot be used to directly view nanometer scale structures, such as the filaments and myosin heads within the sarcomere. Fluorescent and confocal microscopes have achieved the speeds and resolutions required for such analysis, but submicron imaging techniques have largely stalled at less than 100μm from the surface[19,20]. This has led to recent creative techniques such as whispering gallery mode microlasing[15], but this still requires the invasive sample preparation of an inserted microbead, it only measure structures within 200nm of the bead, and the signal-to-background ratio is near 3dB at depths greater than 200μm. As such, present understanding of structural maturation in hiPSC-CMs (**Figure 1a**)[8,18] has been largely developed from 2D models and the aforementioned techniques, where information at depth or in live tissue is difficult to achieve. Alternative approaches to probe subcellular structure in 3D models will provide critical information, tying structure to function, in the modeling and maturation of hiPSC-CMs.

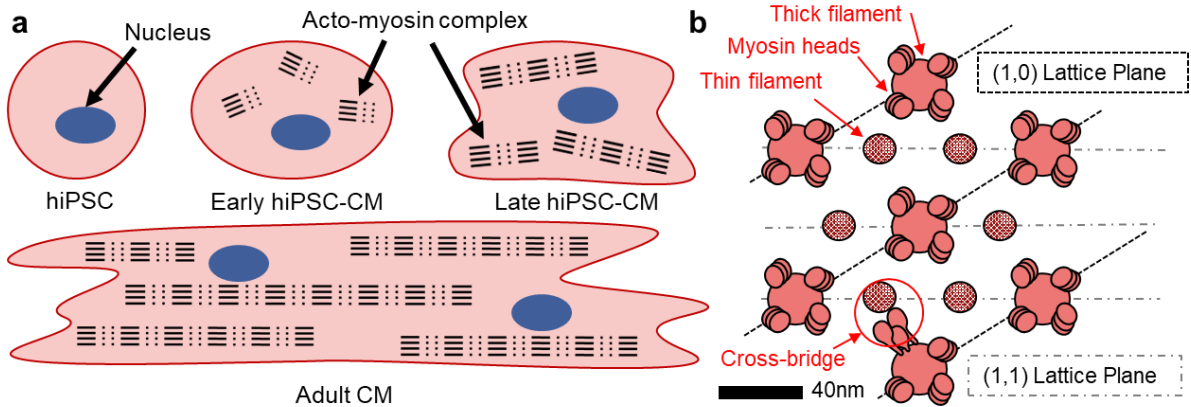

**Figure 1.** Acto-myosin model in cardiac tissue. (a) Structural maturation of sarcomere in hiPSC-CMs (adopted from [22]). (b) Ideal cross-section of acto-myosin complex within the sarcomere of a mature CM (as in adult cardiac muscle). For scale, the shortest spacing between actin filaments is typically 37-45nm in vertebrate cardiac muscle[4,5].

Small angle X-ray scattering (SAXS) is established as a powerful tool for probing nanoscale structure in highly ordered materials and is capable of measuring live tissue dynamics at depth in adult cardiac muscle[21,22] (**Figure 1a**, bottom). In myofilaments, which comprise the primary contractile units within cardiomyocytes, acto-myosin complexes form a hexagonal lattice of filaments that can be detected by SAXS (**Figure 1b**). Scattering intensities of distinct planes relative to myosin and actin provide a direct measure of myofilament lattice order and spacing, which are modulated by the contractile state of the muscle[21-27]. The (1,0) plane can be defined to contain thick myofilaments, while the (1,1) plane consists of mainly thin actin filaments. During contraction, myosin heads extend off the myofilament to form cross-bridges, and mass is transferred from the (1,0) plane to the (1,1) plane. A ratio between scattering intensities of these two planes is analyzed, where the (1,0) intensity is typically much larger than the (1,1) intensity in adult cardiac muscle[21-27]. Despite the advances in cardiac SAXS techniques and the functional maturation in engineered tissue platforms, hiPSC-CMs have not yet been shown to have a level of nanoscale order detectable by SAXS[27]. Noteworthy is that tissues grown in these 3D platforms still exhibit at least a factor of ten-fold lower contractile force than adult cardiac muscle. Given that alignment and striation at the cellular scale has been shown to improve contractile force[1,2], this shortfall in functional maturation may be due to misalignment of the subcellular structure, which largely comprises the myofilament lattice. As such, we refer to the improvement in myofilament lattice ordering as structural maturation.

Regardless of tissue scale or dimension (2D and 3D), electrophysiology[28] and contractile force (CF) generation[1-3] are the primary metrics of functional maturity in engineered heart tissues. The ability to assess the nanoscale structure of engineered tissues would add an important approach to characterizing structural maturation. Here, we leverage the periodic arrangement of myofilaments in mature hiPSC-CMs to detect structural maturation with SAXS. We develop a robust, batch-process methodology to conduct measurements on a high-throughput 3D tissue platform. Both myofilament scattering intensity and lattice spacing are shown to correlate with contractile force in hiPSC-CMs. We show that these new techniques are capable of detecting significant structural differences arising from a myofilament mutation in a genetic model for hypertrophic cardiomyopathy (HCM), compared to a wild type control in hiPSC-CMs. This work highlights the utility of 3D tissue platforms with an engineered mechanical environment and the necessity for methodologies to probe maturation in such advanced 3D tissues.

RESULTS AND DISCUSSION

The primary objectives of this study were to develop a robust methodology for probing the nanostructure of hiPSC-CMs in a 3D environment, to elucidate the structures contributing to the signal, to validate identified trends with SAXS results of explanted cardiac tissue, and to apply the newly developed technique to a genetic model in hiPSC-CMs. For this to be successful, a platform with the ability to detect both functional and structural properties is necessary at a moderately high throughput to generate statistics. Therefore, the microtissue strain gauge (µTUG) platform was selected[10,29], where batches of cardiac microtissues (CMTs) are examined. The hiPSC-CMs are co-cultured with human mesenchymal stem cells and suspended in a 3D fibrin matrix (see Methods). The elastomer pillars are designed with a spherical top to facilitate and maintain tissue attachment. The resulting 3D tissue (approximately 1.2 mm long, 500 µm diameter) compacts for several days and contracts spontaneously in the passive mechanical environment (**Figure 2b**). Deflections of the elastomer pillars are tracked by a microscope and provide a functional readout of peak tissue contractile force. All presented samples exhibit contractile forces within the reported hiPSC-CMT range[1,2].

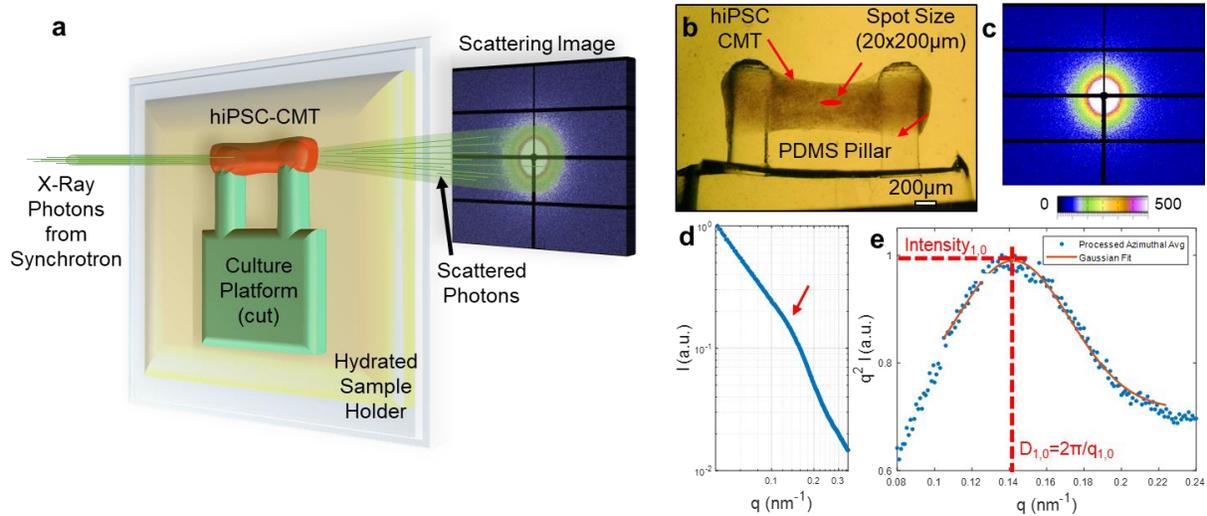

**Figure 2.** SAXS in hiPSC-CMs. (a) Schematic of synchrotron X-ray scattering of 3D tissue on passive mechanical platform. (b) Platform with hiPSC-CMs grown in a 3D microtissue environment. (c) Typical SAXS image from center of tissue. (d) Azimuthal average of scattering image in (c). (e) $q^2$ weighted peak in (d) fit by a Gaussian.

The contractile apparatus of hiPSC-CMs is expected to be loosely organized during early stages of cardiac maturation (illustrated in **Figure 1a**)[8,9]. Therefore, in order to detect structure, samples are probed by synchrotron SAXS, where a high-intensity beam of X-ray photons is transmitted through the sample and scatters elastically off of nanoscale structures at small angles, producing a scattering image recorded with a 2D detector (**Figure 2a**). In this work, the measurements were done at the SMI beamline, NSLS-II, Brookhaven National Laboratory. It is noteworthy that scattering experiments have been conducted previously on a similar tissue system, but the level of hiPSC-CM maturation was determined to be insufficient to discern a signal from the underlying structure[27]. In our experiment, tissue function is recorded first on live tissue and then SAXS is conducted on the same fixed samples at room temperature. During the SAXS measurement, CMTs remain on the posts, hydrated by a buffer solution (see Methods). A spot size of 20 μm x 200 μm in the center of the tissue (**Figure 2b**) is irradiated, producing a scattering image (**Figure 2c**).

To ensure a robust methodology, a customized routine is developed to process the pixel intensities in the scattering images. A circular average of the representative sample (**Figure 2b**) is shown after background subtraction (**Figure 2d**). A Kratky plot, where the signal is multiplied by the square of the wavenumber, is commonly used to extract information from a loosely organized sample[30]. Gaussian fit parameters of the intensity profile (**Figure 2e**) are extracted and compared in a batch-process. In the loosely ordered hiPSC-CMs, the peak intensity is assumed to be largely representative of structural order. The center of the peak in inverse space corresponds to the lattice spacing. The full width at half maximum (FWHM, $\Delta q$) is related to the length over which order persists (correlation length) in a sample with homogenous lattice strain.

The anatomical source of scattering is discerned from the wavenumber (q), literature discussing cardiac SAXS[21-27], and a series of experiments varying matrix and cell conditions. The wavenumber in the representative sample (**Figure 2d**) is $q=0.14 nm^{-1}$ and corresponds to a spacing of $D=44 nm$. This is in the range of typical spacing found for a myofilament lattice in mature tissue[21,24,26,27]. To verify the attribution of the $q=0.14 nm^{-1}$ peak to the myofilament lattice, we treated tissues with collagenase or trypsin, enzymes that digest components of the extracellular matrix (ECM). If the observed scattering peak were due to ordering in the ECM, chemical distortion or degradation of the ECM would affect or diminish that peak. However, we found that peak was preserved in enzyme treated tissues (**Figure S1**). Furthermore, we found that tissues with only non-myocytes as well as decellularized hiPSC-CM tissues did not exhibit a peak near this wavenumber (**Figure S1**). This reaffirms the assertion that the peak at $q=0.14 nm^{-1}$ is representative of the myofilament lattice.

Further information may be extracted from the data regions outside of the peak at $q=0.14 nm^{-1}$. As mentioned earlier, the structural order is assumed to be loosely organized[30]. In order to assess structural geometry of a loosely organized lattice, the sample is assumed to comprise elongated nanoscale structures[31]. Form factors are then extracted from the Porod plot (**Figure 2d**) in the low and intermediate Guinier regions (**Figure S2**) to find an average cylindrical geometry of 22nm radius and 60nm length. Previous work using electron microscopy have found diameters of individual myofilament diameters to be 30-40 nm and actin filaments to be 10 nm[32]. This average geometry over the scattering volume may be suggestive of fragmented fibers or segmented contractile apparatuses in maturing hiPSC-CMs (as illustrated in **Figure 1a**)[8].

An analysis of the myofilament lattice in hiPSC-CMs, compared to adult cardiac tissue, provides a roadmap for structural maturation. Since the X-ray beam passes through the entire tissue, this technique reports a volumetric average with contributions from multiple cells. The myofilament lattice spacing, $D_{1,0}$, of a representative sample (**Figure 2b**) is found to be 44nm and the FWHM is $\Delta q_{1,0}=0.07 nm^{-1}$, or a correlation length of about 100nm. Compared to adult cardiac muscle, where $D_{1,0}$ has been reported near 39nm[23-25], hiPSC-CMs exhibited a slightly larger $D_{1,0}$ and a shorter correlation length. The assertion that contractile units are ordered over a smaller region compared to adult tissue may give a structural perspective to the reduced contractile function seen in hiPSC-CMs.

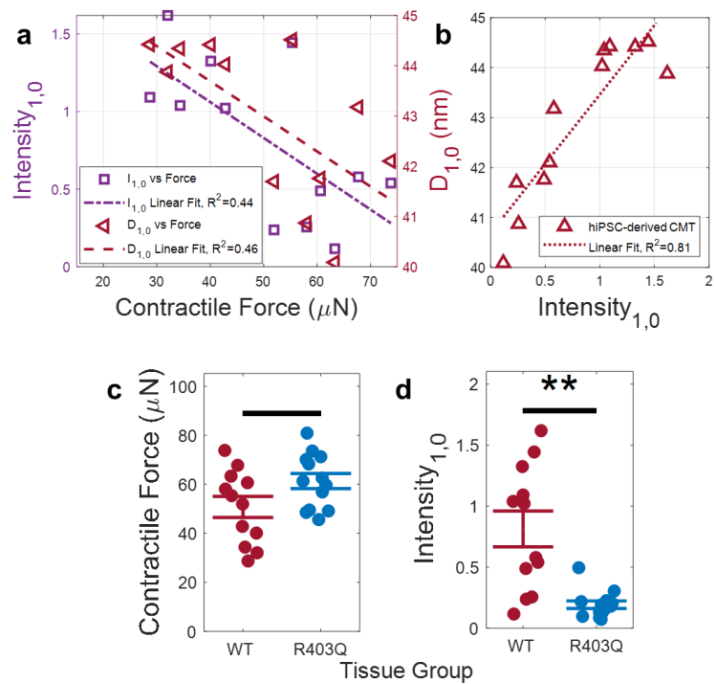

**Figure 3.** Scattering results in hiPSC-CMs. (a) Negative correlation between CF and $I_{1,0}$ as well as $D_{1,0}$ (pCa 2.7). (b) Positive correlation between both SAXS parameters in (a). (c) CF of tissue groups, WT and R403Q$^{+/-}$ (p=0.058). (d) $I_{1,0}$ of tissue groups, where the trend varies inversely to CF in (c), suggesting more cross-bridge formation in R403Q$^{+/-}$ hiPSC-CMs.

With both structural and functional information from the same hiPSC-CMs, we now directly compare the SAXS information with contractile force. To accomplish this, we simply calculate the correlation coefficient between the structural and functional datasets. Here the buffer contains a calcium concentration of pCa 2.7. $D_{1,0}$ has a negative relationship with CF ($R^2$=0.46), suggesting that a tissue capable of stronger CF has a more compact lattice (**Figure 3a**). A negative relationship with (1,0) scattering intensity, $I_{1,0}$, ($R^2$=0.44) suggests that more cross-bridges were formed in a tissue with higher CF. Furthermore, there was a stronger correlation between $I_{1,0}$ and $D_{1,0}$ ($R^2$=0.81), suggesting greater variability in CF measurements than in $I_{1,0}$ and $D_{1,0}$ (**Figure 3b**). This may be explained by the reduced sampling of SAXS to a spot in the center of the CMT compared to the CF of an entire CMT.

To understand how disease-associated myofilament mutations might impact lattice organization, we compared the HCM-causing myofilament mutation, R403Q$^{+/-}$, with an isogenic wild-type (WT) control hiPSC line. The CF of R403Q$^{+/-}$ CMTs was slightly higher than WT (p=0.058) (**Figure 3c**). The CF of a CMT is dependent on the number of hiPSC-CMs and nonmyocytes in the tissue, creating variability, but significantly higher forces have been observed in single hiPSC-CMs of the same cell lines previously[5]. Conversely, the mean $I_{1,0}$ for R403Q$^{+/-}$ hiPSC-CMs was found to be significantly lower than the WT control (**Figure 3d**). This suggests more cross-bridges were formed in R403Q$^{+/-}$ hiPSC-CMs compared to WT, consistent with prior work in adult cardiac muscle[24].

We also analyzed the effect of a relaxation buffer on the contractile structure (**Figure 4**). Relaxation is known to causes myosin heads to aggregate toward the myofilament in adult cardiac muscle[21-27], and so it is expected that relaxation of the contractile apparatus will increase scattering intensity of the myofilament lattice at (1,0). Hypercontractility in tissues with the $R403Q^{+/-}$ mutation may be explained by a higher activity of the myosin heads, extending away from the myofilament and leading to hypercontractility[33]. Therefore, we chose to examine the effect of a relaxation buffer on $R403Q^{+/-}$ CMTs. We observed that $I_{1,0}$ increased in several relaxed samples, consistent with our hypothesis. Furthermore, the distribution of $D_{1,0}$ for those higher intensity samples ($I_{1,0}>0.5$) significantly narrowed, suggesting an improved ordering of the nanoscale structure when in a relaxed state.

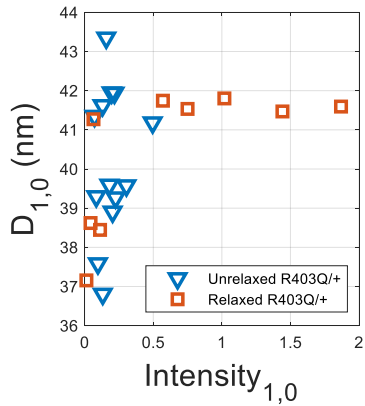

**Figure 4.** Effect of relaxation buffer in hiPSC-CMs. The effect of a 50mM K+ relaxation buffer on tissues with the myofilament mutation, producing higher intensity and more consistent $D_{1,0}$.

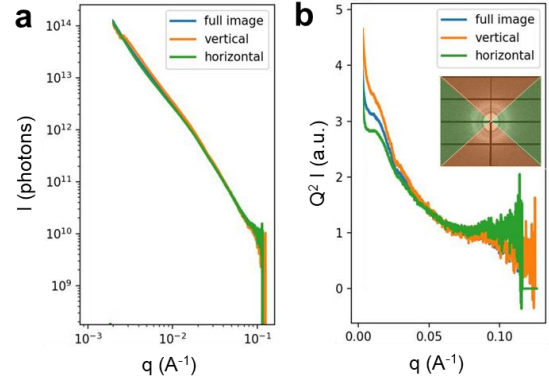

**Figure 5.** Anisotropic Structure in hiPSC-CMs grown in 3D CMTs. Circular averages of scattering image in Figure 2 showing Porod plot (left) and enhanced peaks in Kratky plot (right). Vertical and horizontal denote circular average of only the regions coordinated by color in the inset.

It is noteworthy that, while the scattering has been observed to be largely isotropic (**Figure 2c**), a small degree of anisotropy can be detected (**Figure 5**). The circular average (left) and $q^2$-weighted intensity before normalization (right) are shown. The scattering image is parsed into scattering in the vertical (perpendicular to tissue in **Figure 2b**) and horizontal (along tissue in **Figure 2b**) directions. Filaments oriented in the horizontal direction will scatter vertically and orientation in the vertical direction will scatter horizontally. As tissues predominantly are scattering in the vertical direction, the myofilament lattice is primarily ordered along the tissue, which allows it to produce net contraction along a single-axis, deflecting the pillars.

A 3D tissue environment may be heterogenous, where the thickness of the sample, cardiomyocyte density, and cardiomyocyte nanostructure vary with position. The results presented in **Figures 2-5** are gathered from a single spot in the center of a cardiac tissue, where the thickness and cardiomyocyte density is assumed to be maximum. To understand the heterogeneity (and thickness dependence) within a cardiac tissue, we vary the scattering location, shown in **Figure 6**. Since the intensity is scaled by the low-q region during batch-processing analysis, the tissue is effectively normalized. This is a common scattering technique for low signal-to-noise, hydrated biological samples[30]. The area under the curve of the peak is called the Porod invariant, which is a concentration (or volume) independent value. After scaling, the shape of the peak is expected to be equivalent for thescattering taken at different positions in the tissue. The scattering peaks in **Figure 6** have consistent overlap, suggesting that structure is relatively consistent in a given tissue sample. Additionally, the nanostructural order is far more similar within a given tissue sample, compared to sample-to-sample changes, as is reported in **Figures 3** and **4**, where both peak amplitude $I_{1,0}$ and peak width $D_{1,0}$ are changing. While SAXS averages the scattering of all cells in the volume of the scattered tissue (in our case, an approximate volume of 0.002 mm$^3$), any changes in cell density do not appear to contribute to large variations within a sample. Therefore, despite the heterogeneous 3D tissue environment, we find that our technique and analysis approach is sufficient for nanostructural analysis of the myofilament lattice in hiPSC-CMs.

We hypothesize that there is a threshold of maturation for myofilament detection, as samples exhibiting approximately less than 20μN contractile force did not show adequate scattering signal (and were therefore excluded). Despite the ability of the hiPSC-CMs to functionally contract the tissue, the actin filament lattice peak (1,1) was not observed in the scattering signal. In typical cardiac SAXS analysis of mature tissue, a ratio of the myofilament to actin filament lattice peak intensities ($I_{1,1}/I_{1,0}$) is reported, where the (1,0) signal is much stronger. Therefore, (1,1) signal may be obscured by noise or by the large width of the (1,0) signal peak.

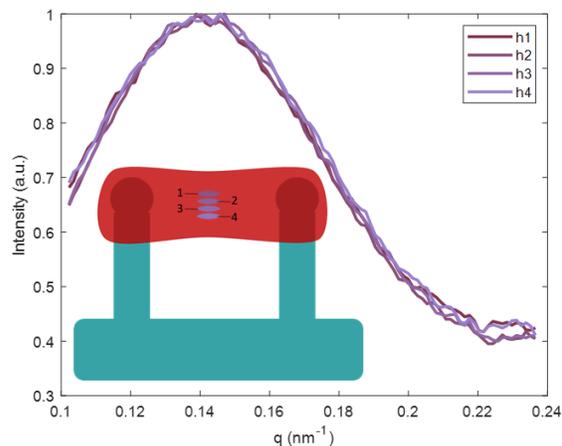

**Figure 6.** Effect of Scattering Location. Within a 3D cardiac tissue, cell density, tissue thickness, and cardiac nanostructure may vary. The Porod invariant (area under peak) is consistent after a concentration (volume) correction in the batch-process. The peak shape is consistent at different spots in the sample, suggesting that nanostructural order is similar within a given sample.

CONCLUSION

Presently, a large part of pharmaceutical research still relies on expensive and inaccurate animal models. Engineered human cardiac tissue will not only provide higher relevance to human studies, but has potential for patient-specific and heart-on-a-chip trials. For this, a greater maturity of cardiac cells is required, and has been improved by multicellular, 3D tissue platforms. Methods for live measurement of cardiac dynamics and maturation in such a 3D environment are lacking. Electron microscopy is very useful, but requires highly invasive preparation in a vacuum environment. Fluorescent and confocal microscopes have achieved the speeds and resolutions required for analyzing many contractile properties, but techniques have largely stalled at less than 100μm from the surface of a 3D tissue. Alternatively, SAXS has been shown as a promising method for nanostructural measurement of live tissue dynamics and contractile properties, and our work is the first step in that direction for engineered cardiac tissue. Here, we demonstrate that engineered cardiac cells, grown from hiPSCs in a 3D tissue environment, comprise internal nanostructure exhibiting periodic spatial arrangement of a high degree. We conclude the observation of the myofilament lattice from the peak position in inverse space, and we discuss the effects of tissue preparation, structural anisotropy, and sample heterogeneity on scattering results. Compared to adult cardiac muscle, we found that hiPSC-CMs exhibit diminished nanoscale ordering but follow relationships with contractile force, the HCM-causing R403Q$^{+/-}$ mutation, and a relaxation buffer. While this is an indication of hiPSC-CM maturation in 3D constructs, recent advances in hiPSC-CM maturation using electromechanical stimulation are ideal candidates for further study.

MATERIALS AND METHODS

For generation and recording of hiPSC-derived CMTs, hiPSCs from the PGP1 parent line and CRISPR-Cas9 PGP1 edited cells with a heterozygous R403Q mutation in the β-myosin heavy chain (MYH7) were received from the Seidman Lab[5]. hiPSCs were maintained in mTeSR1 (StemCell) on Matrigel (Fisher) mixed 1:80 in DMEM/F-12 (Fisher) and split using Accutase (Fisher) at 60-90% confluence. iPSCs were differentiated into iPSC-CMs by small-molecule, monolayer-based manipulation of the Wnt-signaling pathway[34]. Once cells were beating, iPSC-CMs were purified using RPMI, no-glucose media (Fisher) with 4 mM Sodium DL Lactate solution (Sigma) for 2-5 days. Following selection, cells were re-plated and maintained in RPMI with 1:50 B-27 Supplement (Fisher) on 10ug/mL Fibronectin (Fisher) coated plates until day 30+.

CMT devices with tissue wells, each containing two cylindrical micro-pillars with spherical caps, were cast in PDMS from a 3D-printed mold (Protolabs), similar to a previous design[10]. Devices were plasma treated for 60s, treated with 0.01% PLL (ScienCell) for 2hrs, 0.1% Glutaraldehyde (EMS) for 15min, washed 3x with DI water, and let sit in DI water at 4C overnight. Immediately prior to seeding, devices were soaked in 70% ethanol for 30min, dried, and UV-sterilized for 15min. Next, 3µL of 5% BSA was added to the bottom of each tissue well and devices were centrifuged at 3000rpm for 1.5min. After 1hr incubation at RT, BSA was removed and 2µL of 2% Pluronic F-127 (Sigma) was added to each well and incubated for 30min at RT to prevent CMTs sticking to the bottom surface of the devices.

A total of 60,000 cells per CMT, 90% iPSC-CMs and 10% human mesenchymal stem cells (hMSCs), were mixed in 7.5µL of an ECM solution, 4 mg/mL of human fibrinogen (Sigma), 10% Matrigel (Corning) and 1.6 mg/mL Thrombin (Sigma), with 5 µM Y-27632 (Tocris), and 33 µg/mL Aprotinin (Sigma). The cell-ECM mixture was pipetted into each well, and after polymerization for 5 min, in a growth media containing high-glucose DMEM (Fisher) supplemented with 10% Fetal Bovine Serum (Sigma), 1% Penicillin Streptomycin (Fisher), 1% Non-essential Amino Acids (Fisher), 1% Glutamax (Fisher), 5 µM Y-27632, and 33 µg/mL Aprotinin was added and replaced every other day. Y-27632 was removed 2 days after seeding and Aprotinin was decreased to 16µg/mL after 7 days. Time-lapse videos of CMT contraction were then acquired at 30fps using a 4x objective on a Nikon Eclipse Ti with an Evolve EMCCD Camera (Photometrics), equipped with an environmentally controlled chamber. CF measurements were calculated from pillar deflection and the measured stiffness of the pillar (2.67N/m)[29]. After recording, CMTs were fixed in 4% PFA (Fisher) for 30min, washed 3x with PBS, and stored in PBS until SAXS radiation. CMTs fixed in a relaxation buffer were placed in growth media with 50mM KCl for 1hr and fixed in 4% PFA in growth media with 50 mM KCl for 30min.

For small angle X-ray scattering and associated data processing, tissues are placed between 70µm thick glass and kapton tape, remaining stretched on the pillars and in a PBS solution at room temperature. The x-ray wavelength was 0.077nm, the sample-to-detector distance was 8m, the flux was $1\times10^{12}$ ph/s and samples were irradiated for 1.5s at a spot size of 20x200 µm$^2$. This results in a calculated radiation dosage of 0.15 MGy, which is over a factor of 20 lower than previous work with hydrated cardiac cells[27,35] (further detail in Supplementary Material). SAXS images were recorded by a Pilatus3 X 1M detector and batch-processed in Python-based Jupyter notebook, using a combination of data processing techniques. First, the scattering from the tissue is circular averaged to extract an intensity versus wavenumber profile. Background data is collected for each sample by selecting a spot where there is only buffer in the chamber (no tissue), and subtracting from experimental curves. Slope subtraction is employed by fitting the intensity decay in regions outside of our peak of interest, and subtracting the fit. Finally, the region of interest is fit by a Gaussian peak, where all the post-processing settings are consistent for all samples in a batch-process. Student's t-test was used to determine significant difference between CMT groups (**$p<0.005$). Error bars are standard error of the mean.

ACKNOWLEDGMENTs

We thank Olga Antipova for experimental advice on diffraction of cardiac tissue. This work was supported by the NSF CELL-MET ERC award no. 1647837 and NSF GRFP (DGE-1840990; J.K.E). This research used the SMI beamline (12-ID) of the NSLSII, a U.S. DOE Office of Science User Facility operated by Brookhaven National Laboratory under Contract No. DESC0012704.

AUTHOR CONTRIBUTIONS

The manuscript was written through contributions of all authors. All authors have given approval to the final version of the manuscript. J.J., P.C., J.E., A.C., K.L., designed research; J.S. and C.S. developed CRISPR-Cas9 edited cell line, R403Q+/-. P.C., J.E. cultured tissues and recorded contractile force; R.L. designed CMT platform. J.J., K.L., M.Z., G.F. performed scattering and analyzed data; J.S., C.S., K.L., M.Z., C.C., D.B. advised research. J.J. wrote paper with input from all parties.